\documentclass[12pt,showpacs,preprintnumbers]{revtex4}

\usepackage{graphicx}
\usepackage{dcolumn}
\usepackage{bm}
\usepackage{amssymb}
\usepackage{epsfig}    
\def\beq{\begin{equation}}
\def\eeq{\end{equation}}
\def\ba{\begin{array}}
\def\ea{\end{array}}
\def\bea{\begin{eqnarray}}
\def\eea{\end{eqnarray}}

\def\sq2{\sqrt{2}}
\def\scrs{\scriptstyle}
\def\End{\end{document}}

\begin{document}

\title{Electromagnetic dipole moments of the top quark}%
\author{%
{Antonio~O.~Bouzas}~~and~~{F.~Larios}\footnote{
  larios@mda.cinvestav.mx, corresponding author.}}
\affiliation{%
\vspace*{2mm} 
Departamento de F\'{\i}sica Aplicada,
CINVESTAV-M\'erida, A.P. 73, 97310 M\'erida, Yucat\'an, M\'exico
}

\begin{abstract}
\hspace*{-0.35cm}

Recent measurements like the $t\overline{t}\gamma$ production by CDF
as well as the Br($B\to X_s \gamma$) and $A_{CP}$($B\to X_{s} \gamma$)
are used to constrain the magnetic and electric dipole moments of the
top quark.  The $B\to X_{s} \gamma$ measurements by themselves define
an allowed parameter region that sets up stringent constraints on both
dipole moments.  Actually, significantly more stringent than
previously reported.  The measurement by CDF has a $\sim37 \%$ error
that is too large to set any competitive bounds, for which a much
lower $5\%$ error would be required at least.  On the other hand,
because of the LHC's higher energy (apart from its higher luminosity)
the same measurement performed there could indeed further constrain
the allowed parameter region given by the $B\to X_{s} \gamma$
measurement.


\end{abstract}

\maketitle

\setcounter{footnote}{0}
\renewcommand{\thefootnote}{\arabic{footnote}}

\section{Introduction}

The top quark stands out as the heaviest known elementary particle and
its properties and interactions are among the most important
measurements for present and future high energy colliders
\cite{topreviews}.  In particular, anomalous top dipole moments could
point towards new physics (henceforth NP) effects like a composite
nature of the top quark \cite{tait}.
Concerning the anomalous magnetic and electric dipole moments
(henceforth MDM and EDM, respectively) of the top quark, it is well
known that the Br($B\to X_s \gamma$) can set up the most stringent
constraints \cite{kamenik}.
We will make a re-evaluation of those constraints, where in addition
to the Br($B\to X_s \gamma$) we will consider a CP asymmetry
for this process that indeed sets the strongest bounds
on the EDM of the top quark.  As we shall see, our bounds are
more stringent than reported previously, and they are consistent
with bounds that can be (indirectly) inferred from other studies.
Recently, it has been suggested that another possible test of the
MDM of the top quark could come from $H\to \gamma \gamma$
\cite{rafelski}.  However, until such rare decay process gets
more precise experimental analysis this option will not be
feasible, and $b\to s \gamma$ along with $t\bar t \gamma$
production will remain the best probes of the MDM and EDM
of the top quark.
The CDF collaboration has reported a measurement of $t\bar t
\gamma$ production with $6 fb^{-1}$ of data \cite{cdfmeasurement}.
(Some preliminary study has also been done for the LHC
\cite{atlasthesis}.)  This process has been considered as a probe of
the dipole moments of the top quark by Baur et. al \cite{baur04} and
their overall conclusion was that even though the Tevatron would not
be able to set bounds as stringent as those from the $B\to X_s \gamma$
measurements, the LHC could.  The reason for this is that since the
dipole coupling is proportional to the momentum of the photon there is
more relative contribution (compared to the QED coupling) as the
energy of the collider increases.  In this work, we take the
experimental result by \cite{cdfmeasurement} and make an estimate of
the bounds on the MDM and EDM, where indeed we corroborate that $t\bar
t \gamma$ at the Tevatron is far from competing with $B\to X_s
\gamma$.  But on the other hand, we also find that the LHC could in
principle set significant \emph{direct} bounds that would further
improve what we already have from the \emph{indirect} bounds from
$B\to X_s \gamma$.


\section{The MDM and EDM of the Top quark: previous studies.}

Following \cite{hewett}, we define the effective
$t\overline{t}\gamma$ Lagrangian 
\bea
{\cal L}_{t\overline{t}\gamma} = e \overline{t} \left( Q_t \gamma_\mu
  A^\mu +  
\frac{1}{4m_t} \sigma_{\mu\nu} F^{\mu \nu}
(\kappa + i\widetilde{\kappa} \gamma_5) \right) t,
\label{dipoledefinition}
\eea
where the CP even $\kappa$ and CP odd $\widetilde{\kappa}$ terms
are related to the anomalous MDM and EDM of the top quark, respectively.
Similar Lagrangians are also defined in
\cite{kamenik} and \cite{baur04}.  Comparing their 
different notations (notice a relative minus sign in the charge
term) we obtain the following relations,
\bea
\kappa &=& -F^\gamma_{2V} \; =\; \frac{2m_t}{e} \mu_t \; =\;
Q_t a_t \; ,\nonumber \\
\widetilde \kappa &=& F^\gamma_{2A} \; =\; \frac{2m_t}{e} d_t
\; , \label{definitions}
\eea
where $a_t = (g_t-2)/2$ is the anomalous MDM in terms of the
gyromagnetic factor $g_t$.  The factors $F^\gamma_{2V}$ and
$F^\gamma_{2A}$ are used in \cite{baur04} and 
$\mu_t$ and $d_t$ in \cite{kamenik}.
The SM prediction for $a_t$ is $a_t^{\rm SM} =0.02$
\cite{bernreuther}, which translates to $\kappa^{\rm SM} = 0.013$.
The bounds for $\kappa$ that we will obtain will be about two
orders of magnitude greater, therefore the SM prediction will not
be considered in our calculations.
On the other hand, the CP violating EDM factor $d_t$ is strongly
suppresed in the SM: $d^{\rm SM}_t < 10^{-30} e\, $cm
($\widetilde \kappa < 1.75\times 10^{-14}$)\cite{smedm}.
The EDM is thus a very good probe of new physics.  There are
models with vector like multiplets that predict values as high
as $10^{-19} e\, $cm ($\widetilde \kappa < 1.75\times 10^{-3}$)
\cite{nathelectric}.  In fact, these models can also predict
large values of other CP odd top quark properties like
the chromoelectric dipole moment \cite{nathchromo}.
There are bounds based on the indirect effects on the EDM of
the neutron, it has been found that
$d_t < 3\times 10^{-15}$ ($\widetilde \kappa < 5.25\times 10^{1}$)
\cite{toscano}.  This is a rather weak bound compared to the
ones we find below based on the branching ratio and the
CP assymetry of $b\to s \gamma$.
As mentioned before, in~\cite{baur04} a study is made on the
sensitivity of the Tevatron and the LHC to measure $\kappa$
and $\widetilde \kappa$ through $t\bar t \gamma$ production.
Their conclusion for the Tevatron (with $8 {\rm fb}^{-1}$
of integrated luminosity) was that both coefficients could
be probed in the range $\pm 5.2$ at $68.3 \% CL$.  For the
LHC at $\sqrt{s} =14$TeV the range would be about $\pm 0.2$
assuming a $300 {\rm fb}^{-1}$ of data.  As we shall see below
those numbers are consistent with our conclusions, even though
our strategy based on the $\sigma_{tt\gamma}/\sigma_{tt}$
ratio is different from the one used in \cite{baur04}.

The MDM and the EDM of the top quark have also been studied in
the context of the $SU(2)\times U(1)$ gauge invariant effective
Lagrangian \cite{buchmuller}.  For instance, a recent
study on the ILC potential to probe the $tt\gamma$ and $ttZ$
vertices can be found in~\cite{fiolhais}.  That study was
made in the context of a minimal list of independent
operators that give rise to couplings involving the top
quark \cite{Aguilar-Saavedra}.  Indeed, the original list by
Buchmuller and Wyler contains a long list of gauge invariant
operators that were supposed to be independent \cite{buchmuller}.
It was found out some years later that some of the operators
involving the top quark were in fact redundant \cite{wudka}.
A recent in-depth analysis made by
Aguilar-Saavedra~\cite{Aguilar-Saavedra} has yielded a short list of  
only eight operators. More recently, a revised general list of all
gauge invatiant operators not necessarily related to the top
quark was given in~\cite{rosiek}.
There are two, and only two, operators that give rise to
both, the MDM and the EDM of the top quark \cite{fiolhais},
\bea
{\cal O}_{uB\phi}^{33} &=& C_{uB\phi}^{33} \,
\bar q_{L3} \sigma^{\mu \nu} t_R \widetilde \phi B_{\mu \nu}
\; + h.c. \; ,\label{oub} \\
{\cal O}_{uW}^{33} &=& C_{uW}^{33} \,
\bar q_{L3} \sigma^{\mu \nu} \tau^a t_R \widetilde \phi
W^a_{\mu \nu} \; + h.c. \nonumber
\eea
Comparing with the effective Lagrangian used in \cite{fiolhais}, we
obtain the 
relations $d_V^\gamma=-\kappa/2$ and $d_A^\gamma = -\widetilde \kappa/2$.
Then, from Eq.~(2.5) of \cite{fiolhais}:
\bea
\kappa &=& -\frac{2\sq2}{e} \frac{vm_t}{\Lambda^2}
{\rm Re} [ s_w C_{uW}^{33} + c_w C_{uB\phi}^{33} ]
\nonumber \\
\widetilde \kappa &=& -\frac{2\sq2}{e} \frac{vm_t}{\Lambda^2}
{\rm Im} [ s_w C_{uW}^{33} + c_w C_{uB\phi}^{33} ]
\nonumber
\eea
Concerning the contribution from ${\cal O}_{uW}^{33}$ the
ATLAS collaboration has already set bounds on the real
part of the coefficient \cite{fiolhais, atlasw}, 
$-1 < \Lambda^{-2} {\rm Re} [ C_{uW}^{33} ] < 0.5 {\rm TeV}^{-2}$.
Moreover, we can also find a recent similar bound based
on precision electroweak measurements \cite{willenbrock},
$-1.6 < \Lambda^{-2} {\rm Re} [ C_{uW}^{33} ] < 0.8 {\rm TeV}^{-2}$.
This means that $\kappa$ could only reach values of order
$0.2$ coming from this operator.  We shall therefore
ignore the effects from ${\cal O}_{uW}^{33}$ and instead
focus our attention on the operator ${\cal O}_{uB\phi}^{33}$.
The contribution from this operator to the $b\to s \gamma$
process would indeed enter via the MDM and EDM terms of
the $tt\gamma$ vertex (in the unitary gauge) applied inside the
loop associated to the $C_7$ Wilson coefficient \cite{hewett}.
From~\cite{willenbrock} we can find a recent bound on
$ C_{uB\phi}^{33}$,
$-0.5 < \Lambda^{-2} {\rm Re} [ C_{uB\phi}^{33} ] < 10.1 {\rm TeV}^{-2}$.
From the relation
$\kappa = -0.34 {\rm TeV}^2 {\rm Re} [C_{uB\phi}^{33}] \Lambda^{-2}$ 
we conclude that the contribution from this operator should be
in the range $-3.4 < \kappa < 0.17$.  This range is similar
to the limits we have found based on $b\to s \gamma$.

\section{Limits from $t\overline{t}\gamma$ production at the Tevatron}
\label{sec:ttg}

The CDF collaboration has reported a cross--section measurement of
top--quark pair production with an additional photon that carries at
least 10 GeV of transverse energy, $\sigma_{t\overline{t} \gamma} =
0.18 \pm 0.08$ pb \cite{cdfmeasurement}.  In addition, using events
with the same selection criteria as for the $t\overline{t} \gamma$
candidates, but without the photon, they also perform a measurement of
the $t\overline{t}$ production cross section.  In this way they
determine the ratio $R^{\rm exp} \equiv
\sigma_{t\overline{t} \gamma}/\sigma_{t\overline{t}} = (0.024 
\pm 0.009)$,
in which systematic uncertainties are
eliminated.  That experimental result is in excellent
agreement with the SM prediction $R^{\rm SM} = 0.024 \pm 0.005$
\cite{cdfmeasurement}.

The potential of using $t\overline{t}\gamma$ production at hadron
colliders as a probe of the $t\overline{t}\gamma$ vertex was studied
in \cite{baur04} (following previous work in \cite{baur01}).  That
production process can probe the charge of the top quark, including
the presence of an axial--vector term, if any.  The strategy proposed
in \cite{baur04} relies on analyzing the transverse--momentum
distribution of the radiated photon, as the $\sigma_{\mu \nu} q^\nu$
dipole term tends to favor a greater $p_T^\gamma$.  In this paper we
assume that the dimension-4 $t\overline{t}\gamma$ coupling is as
dictated by the Standard Model, so that possible NP effects appear in
the dipole terms only.  Since the main result by CDF is given in terms
of $\sigma_{t\overline{t}\gamma}/\sigma_{t\overline{t}}$, we consider
that ratio as a function of the MDM $\kappa$ and the EDM
$\widetilde{\kappa}$ to set bounds on those parameters.  Although our
strategy is simpler than the analysis carried out in
\cite{baur04}, we believe it is yet useful to obtain an estimate of
the sensitivity of the Tevatron result, and of future LHC results.
                                                  
In order to quantify the impact of the top--quark MDM and EDM on the
cross section, we focus our attention on the normalized ratio 
\begin{equation}
  \label{eq:rhat}
\widehat{R} \equiv \frac{R}{R^{\rm SM}} =
\frac{\sigma_{t\overline{t}\gamma}}
{\sigma_{t\overline{t}\gamma}^\mathrm{SM}}.   
\end{equation}
In this way, the CDF result can be translated to
$\widehat R^{\rm exp} = 1 \pm 0.375$.  We compute the cross sections
for $p\overline{p} \to t\overline{t} \to b W^+ \overline{b} W^-\gamma
\to \mathrm{FS}$ at leading order at the Tevatron energy, and the same
processes with $pp$ initial state at LHC energies.  We choose
semileptonic final states $\mathrm{FS}$, as done in the CDF
measurement, but consider also a simplified process with final state
$b W^+ \overline{b} W^-\gamma$ as a cross check of our results.  For the
numerical computation of the semileptonic cross section we consider
the process $pp$, $p\overline{p} \rightarrow t\overline{t} \rightarrow
b \overline{b} q q' \ell \nu_\ell \gamma$ with three lepton flavors,
where the final photon can originate from any initial, intermediate or
final charged particle.  The calculation was carried out with
\textsc{Madgraph} 5 \cite{mgme5}, with the set of default parameters
in which $\alpha$, $\sin\theta_W$ and $G_F$ are the primary
parameters, but with $m_t=173$ GeV.  For the parton distributions
functions we use the set CTEQ 6m for proton and antiproton with fixed
renormalization and factorization scales set to $m_t$.  Although not
reported in detail here, we have explicitly checked that the
dependence of our results on the choice of scale is quite weak.

In the radiative production process two modes are predominant: (1)
$t\overline{t}$ produced along with the radiated photon followed by
the decay of the top pair, which is indeed $t\overline{t}\gamma$
production proper, and (2) $t\overline{t}$ produced on-shell with one
of them decaying radiatively (such as $t\to b W^+ \gamma$).  The first
mode may involve initial--state radiation if the initial partons are
charged.  The second mode may involve final--state radiation from the
$b$ jets, the intermediate $W$ boson or the $W$ decay products.  
At the Tevatron energy $\sqrt{s}=2$ TeV, the production of
$t\overline{t}$ and $t\overline{t}\gamma$ receives its dominant
contribution from $u\overline{u}$ initial states, but we take into
account also the smaller contributions from initial $d\overline{d}$
and $gg$. The corresponding scattering amplitudes with two resonant
top/antitop propagators involve a total of 876 Feynman diagrams, as
given by \textsc{Madgraph}, of which 612 are independent. By analogy
with the measurement reported by CDF \cite{cdfmeasurement}, we apply
cuts in the transverse energy of the photon, missing transverse energy
and pseudorapidity of the final particles given by
\begin{equation}
  \label{eq:cuts}
  E_T^\gamma > 10 \mathrm{GeV},\
\quad
\not\!\!E_T > 20 \mathrm{GeV},
\quad
|\eta_q| < 3.6,
\quad
|\eta_b| < 2,
\quad
|\eta_\gamma| < 1,
\quad
|\eta_\ell| < 1.
\end{equation}
With those cuts we obtain a SM cross section $\sigma_{t\overline
  t\gamma}^\mathrm{SM} = 0.07261$ pb at $\sqrt{s}=2$ TeV, in agreement
with the leading-order result reported in \cite{cdfmeasurement}.  In
order to increase the sensitivity of the process to the dipole moments
of the intermediate top quarks it is necessary to reduce the background
from photons originating in final--state charged particles.  For that
purpose we impose a lower bound on the distance from the photon to the
charged particles in the $\eta$--$\phi$ plane, $\Delta = \sqrt{(\Delta
  \eta)^2 + (\Delta\phi)^2}$,
\begin{equation}
  \label{eq:cuts2}
  \Delta_{\gamma,\mathrm{ch}} > 0.4,
\end{equation}
which plays the same role as the analogous cuts introduced in the
actual measurement \cite{cdfmeasurement}.  With the cuts
(\ref{eq:cuts}), (\ref{eq:cuts2}) the SM cross section at 2 TeV is
$\sigma_{t\overline t\gamma}^\mathrm{SM} = 0.0193$ pb.  We also
perform the same computation for $t\overline{t}\gamma$ production in
$pp$ collisions at the LHC both at $\sqrt{s}=7$ TeV and $\sqrt{s}=14$
TeV. In this case the dominant contribution to the production process
comes from $gg$ initial states, but we also take into account the
smaller contributions from the initial states $u\overline{u}$ and
$d\overline{d}$.  Thus, in particular, the set of Feynman diagrams
involved in the scattering amplitudes is the same as in the previous
case.  With the cuts (\ref{eq:cuts}), (\ref{eq:cuts2}), the SM cross
sections at 7 and 14 TeV are $\sigma_{t\overline t\gamma}^\mathrm{SM}
= 0.1770$ and 0.8034 pb, respectively.

On the theoretical side, it is well known that at tree level the SM
amplitude is real.  The CP--even MDM $\kappa$ term contributes
linearly to the real part of the total amplitude, whereas the CP--odd
EDM $\widetilde{\kappa}$ contributes to its imaginary part only.
Therefore, the expression for $\widehat R$ must have in general the
quadratic form $\widehat R = 1 + a_1 \kappa + a_2 \kappa^2 + b_2
{\widetilde{\kappa}}^2$.  By computing $\widehat R$ for several values
of $\kappa$ and $\widetilde{\kappa}$ we can obtain the coefficients
$a_i$ in $\widehat R$ at the desired energy. Then we use a relation
of the form
\bea 
\widehat R_1 < \widehat R=1+ a_1 \kappa + a_2 \kappa^2 + a_3
{\widetilde{\kappa}}^2 < \widehat R_2 
\label{tevatronratio}
\eea 
to find the allowed parameter region for
($\kappa$,$\widetilde{\kappa}$) at that energy.  In the case of the
CDF measurement, we set $\widehat R_{1,2} = 1\pm
0.375$ to define the allowed region at the 1$\sigma$ level.  The
computation of $\sigma_{t\overline{t}\gamma}$ for different values of
$\kappa$, 
$\widetilde{\kappa}$ was carried out by implementing the
effective Lagrangian (\ref{dipoledefinition}) in \textsc{Madgraph} by
means of the program \textsc{FeynRules} 1.6.11 \cite{feynrul} (see
also \cite{ask12} for a more recent description).  The resulting
numerical coefficients in 
(\ref{tevatronratio}) are given by $a_1=\, -0.002,
\, -0.008, \, -0.009$,  $a_2 =\, 0.011, \, 0.055, \, 0.088$ and
$a_3 =\, 0.011, \, 0.055, \, 0.089$ at the Tevatron and LHC energies:
2, 7 and 14 TeV, respectively.  


\section{Limits from $B\to X_s \gamma$}

In the context of effective lagrangians the $b\to s \gamma$ transition
occurs through the effective Wilson coefficient $C_7 (\mu)$, computed
at the electroweak scale $\mu_h \gtrsim M_W$ from loop diagrams where
the photon can be emitted either from the $W$ boson or from the top
quark \cite{buras93}.  NP effects on $C_7 (\mu_h)$ can come from
several different sources, for instance an anomalous $WW\gamma$
coupling.  In this paper we are interested in the contributions from
the MDM and EDM of the top quark to the effective $t\bar t \gamma$
vertex and, for simplicity, those are the only ones we will
consider.  Furthermore, the QCD running of $C_7 (\mu)$ from the
electroweak scale down to the bottom mass scale causes it to mix with
other coefficients, so that $C_7 (m_b)$ can receive NP contributions
also from non-electroweak anomalous couplings.  The main contribution
of this type comes from the Wilson coefficient $C_8 (\mu_h)$
associated with the $t\overline{t}g$ vertex.  If we separate the SM
value $C_7^{\rm SM} (m_b) = -0.31$ from the NP contributions, the form
of $C_7 (m_b)$ in terms of the Wilson coefficients evaluated at
$\mu_h$ is \cite{buras93} 
\bea 
C_7 (m_b) = -0.31 + 0.67\,\delta C_7(\mu_h) + 
          0.09\,\delta C_8(\mu_h) +\cdots,
\label{c7running}
\eea 
where $\delta C_i=C_i-C_i^{\rm SM}$ and the ellipsis refers to
terms containing other Wilson coefficients that make numerically
smaller contributions.  As mentioned above, we will focus only on the
contributions to $C_7(m_b)$ arising from the MDM and EDM of the top
quark.  Thus, in (\ref{c7running}) we set $\delta C_8(\mu_h)=0$ and 
keep $\delta C_7(\mu_h)$ which is given by \cite{hewett}
\bea
G_2 &=& \frac{1}{4} - \frac{1}{x-1} + \frac{\ln x}{(x-1)^2}
\, = 0.0908 \; , \nonumber \\
G_1 &=& \frac{x/2 -1}{(x-1)^3}
\left( x^2/2 - 2x + 3/2 + {\ln x} \right) - G_2 
\, = 0.0326 \; ,\nonumber \\
C_7 (\mu_h) &=& C_7^{\rm SM}(\mu_h) +
\kappa G_1 + i {\widetilde{\kappa}} G_2 \label{c7kap},
\eea
with $x=(m_t/m_W)^2 = 4.63$.  Notice that in (\ref{c7kap}) $C_7^{\rm
  SM}(\mu_h)=-0.22$ is a real number, as is the CP--even MDM term
proportional to $\kappa$, but the CP--odd EDM term in
$\widetilde{\kappa}$ adds an imaginary part to $C_7$.  This means that
the $b\to s\gamma$ width, being proportional to $|C_7|^2$, will depend
linearly and quadratically on $\kappa$, but only quadratically on
$\widetilde{\kappa}$.  On the other hand, studies that involve $b\to s$
transitions in general have been done that can set bounds on the real
part of $\delta C_7$:
$-0.15 < {\rm Re}(\delta C_7(\mu_h)) < 0.03$  \cite{straub}.
Since from (\ref{c7kap}) we get ${\rm Re}(\delta C_7(\mu_h))
= 0.0326 \kappa$, the allowed region for $\kappa$ would be $-5 <
\kappa < 1$.  This result is consistent with the bounds we obtain
below based on the branching ratio for $B \to X_s \gamma$.

\subsection{Limits from the branching ratio ${\cal B} (B\to X_s
  \gamma)$} 

An updated numerical expression for the branching ratio ${\cal B}(B\to
X_s \gamma)$ in terms of the coefficients $C_7(\mu_h)$ and
$C_8(\mu_h)$ can be found in eq.\ (4.3) of \cite{lunghi06} which,
retaining only LO contributions, can be written as
\bea
\lefteqn{
\delta{\cal B}(B\to X_s \gamma)\equiv 
{\cal B}(B\to X_s \gamma) - {\cal B}^{\rm SM}(B\to X_s \gamma) =
10^{-4} \times}
\label{bflunghi} \\
&\times& \left(\rule{0pt}{12pt}
{\rm Re}(-7.184\,\delta C_7 - 2.225\,\delta C_8 +
2.454\,\delta C_7\,\delta C^*_8)  + 
4.743\,|\delta C_7|^2 + 0.789\,|\delta C_8|^2 
\right),
\nonumber
\eea
where $\delta C_{7,8}$ are defined as in (\ref{c7running}) and it is
understood that they are evaluated at the electroweak scale $\mu_h$.
The numerical coefficients in (\ref{bflunghi}) were computed in
\cite{lunghi06} assuming a cut in the photon energy $E_\gamma > E_0 =
1.6$ GeV, as is conventionally done in this type of calculations and
as will always be assumed in this paper in connection with the process   
$B\to X_s \gamma$.  If the only NP effects we take into account are
the MDM and EDM of the top quark, the coefficent $\delta C_7(\mu_h)$
appearing in (\ref{bflunghi}) is given by (\ref{c7kap}), and $\delta
C_8(\mu_h)=0$. 

In order to use (\ref{bflunghi}) to constrain $\kappa$ and
$\widetilde{\kappa}$ we need a predicted value for ${\cal B}^{\rm
  SM}(B\to X_s \gamma)$ and a measured value for ${\cal B}(B\to X_s
\gamma)$.  For ${\cal B}^{\rm SM} (B\to X_s \gamma)$ there are three
recent calculations referred to in the literature, $10^{4}\times{\cal
  B}^{\rm SM} (B\to X_s \gamma) = (2.98 \pm 0.26)$ \cite{becher},
$(3.15 \pm 0.23)$ \cite{misiak}, and $(3.47 \pm 0.48)$
\cite{andersen}. A thorough discussion of those results can be
found in \cite{paz}.  For concreteness, we use in our calculations the
value from \cite{misiak}.  The most recently updated experimental
value is
${\cal B}^{\rm Exp.}(B\to X_s \gamma) = (3.43 \pm 0.21 \pm 0.07)
\times 10^{-4}$ \cite{hfag}  (see also the recent status report
\cite{hurth}). 
With those theoretical and experimental values, from (\ref{bflunghi})
with $\delta C_7$ as given by (\ref{c7kap}), we get the relation
\bea
10^4\times\delta{\cal B}(B\to X_s \gamma) = 
(3.43 \pm 0.22) - (3.15 \pm 0.23) = 
- 0.234 \kappa + 0.005 \kappa^2  + 0.039 \widetilde{\kappa}^2.
\label{brnum}
\eea
which we use to set limits on $(\kappa,\widetilde{\kappa})$.


\subsection{Limits from the asymmetry $A_{\rm CP} (B\to X_s \gamma)$} 

The CP asymmetry 
\bea
A_{\rm CP} (B\to X_s \gamma) &=&
\frac{\Gamma(\bar B\to X_s \gamma)-\Gamma(B\to X_{\bar s}\gamma)} 
{\Gamma(\bar B\to X_s \gamma)+\Gamma(B\to X_{\bar s}\gamma)} 
\label{acp}
\eea 
was first proposed in \cite{neubertcp98}.  Its latest
experimental value is quoted in \cite{hfag} as $A^{\rm Exp.}_{\rm
  CP}(B\to X_s \gamma)=(-0.8 \pm 2.9)\%$. An expression for the
asymmetry that includes the SM contribution as well as NP effects
entering through the Wilson coefficients $C_i$ with $i=1,7,8$ is given
in \cite{neubertcp11} (see also \cite{paz1212}).  Since we are assuming
$C_1 = C_1^{\rm SM}$ and $C_8 = C_8^{\rm SM}$ we can rewrite that
expression in a simplified form.  Following \cite{neubertcp11} we
define the parameters $r_7$, $\theta_7$ as $r_7
e^{i\theta_7}=C_7(m_b)/C_7^{\rm SM}(m_b)$.  With $C_7(m_b)$ from
(\ref{c7running}) and $\delta C_7(\mu_h)$ from (\ref{c7kap}), they
are found to be given by
\begin{equation}
  \label{eq:r7th7}
r_7 e^{i\theta_7} = 1-0.0705 \kappa -i 0.1962 \widetilde{\kappa}.
\end{equation}
We can then write eq.\ (13) of \cite{neubertcp11} as
\bea 
A_{\rm CP} [\%] &=& (a_7+0.5036 d_7
)\frac{\sin (\theta_7)}{r_7} \, + \, (0.6783+1.1550 d_7) \frac{\cos
  (\theta_7)}{r_7} \, +\,
\frac{0.0302}{r^2_7}\, , \label{acp13}\\
a_7 &=& 16.6858 + 2.1400 \frac{\widetilde{\Lambda}^c_{17}}{10 {\rm
    MeV}} + 3.9933 \frac{\widetilde{\Lambda}_{78}}{100 {\rm MeV}} ,
\qquad d_7 =
\frac{\widetilde{\Lambda}^u_{17}-\widetilde{\Lambda}^c_{17}} {300
  \textrm{MeV}}\, ,\nonumber 
\eea 
where the angle $\gamma$ appearing in \cite{neubertcp11} has been set
to $\gamma = 66.4^\circ$, as done in that reference.  The
dimensionless parameters $a_7$, $d_7$ in (\ref{acp13}) are linear
combinations of the hadronic parameters $\widetilde{\Lambda}^u_{17}$,
$\widetilde{\Lambda}^c_{17}$ and $\widetilde{\Lambda}_{78}$,
introduced in \cite{neubertcp11}, that are related to the contribution
of resolved photons to the asymmetry.  The precise values of those
hadronic parameters are not known, but their expected ranges of
variation are estimated  to be \cite{neubertcp11}, 
$-330<\widetilde{\Lambda}^u_{17}<525$ MeV,
$-9<\widetilde{\Lambda}^c_{17}<11$ MeV and
$17<\widetilde{\Lambda}_{78}<190$ MeV.  Thus, for the parameters
appearing in (\ref{acp13}) we have $15.4387<a_7<26.6271$ and
$-1.1367<d_7<1.7800$.  This means, in particular, that the SM
prediction $A^{\rm SM}_{\rm CP} = \left.A_{\rm CP}\right|_{\kappa=0=
  \widetilde{\kappa}}$ is afflicted by a significant uncertainty,
$-0.6 \% < A^{\rm SM}_{\rm CP} < 2.8 \%$.

We treat our ignorance of the hadronic parameters as a systematic
uncertainty in the theoretical computation.  Thus, we set
\begin{equation}
  \label{eq:hadstat}
  a_7=\overline{a}_7\pm \delta a_7 = 21.0329 \pm 5.5942,
\qquad
  d_7=\overline{d}_7\pm \delta d_7 = 0.3217 \pm 1.4583
\end{equation}
in (\ref{acp13}), to obtain $A_{\rm CP}(\kappa,\widetilde{\kappa})=  
\overline{A}_{\rm CP}(\kappa,\widetilde{\kappa}) \pm 
\delta A_{\rm CP}(\kappa,\widetilde{\kappa})$ with
\bea
\overline{A}_{\rm CP}[\%] &=& 
\left.\rule{0pt}{14 pt}
A_{\rm CP}[\%]\right|
\raisebox{-12 pt}{$
\renewcommand{\arraycolsep}{0 pt}
  \begin{array}[b]{ccc}
\scrs a_7 &\scrs =& \scrs \overline{a}_7\\[-12pt]
\scrs d_7 &\scrs =& \scrs \overline{d}_7
  \end{array}$
}
=\frac{1.0801-0.0740 \kappa - 4.1594 \widetilde{\kappa}}
{(1-0.0705\kappa)^2 + 0.0385\widetilde{\kappa}^2} , \nonumber\\
\delta A_{\rm CP}[\%] &=& 
\frac{1}{r_7} \sqrt{\sin(\theta_7)^2 (\delta a_7)^2 +
(1.1550 \cos(\theta_7) + 0.5036 \sin(\theta_7))^2 
(\delta d_7)^2} \label{means}\\
&=& \frac{\sqrt{2.1267 (1.1550 - 0.0814\kappa - 0.0988 
\widetilde{\kappa})^2 + 1.2053 \widetilde{\kappa}^2}}
{(1-0.0705\kappa)^2 + 0.0385\widetilde{\kappa}^2}
\nonumber.
\eea
With the asymmetry written in this form, we can use its experimentally 
measured value to set bounds on the allowed region for 
$(\kappa,\widetilde{\kappa})$. 


\section{Allowed parameter space for ($\kappa$,$\widetilde{\kappa}$)}
\label{sec:allowed}

\begin{figure}[p]
  \centering
  \scalebox{1.0}{\includegraphics{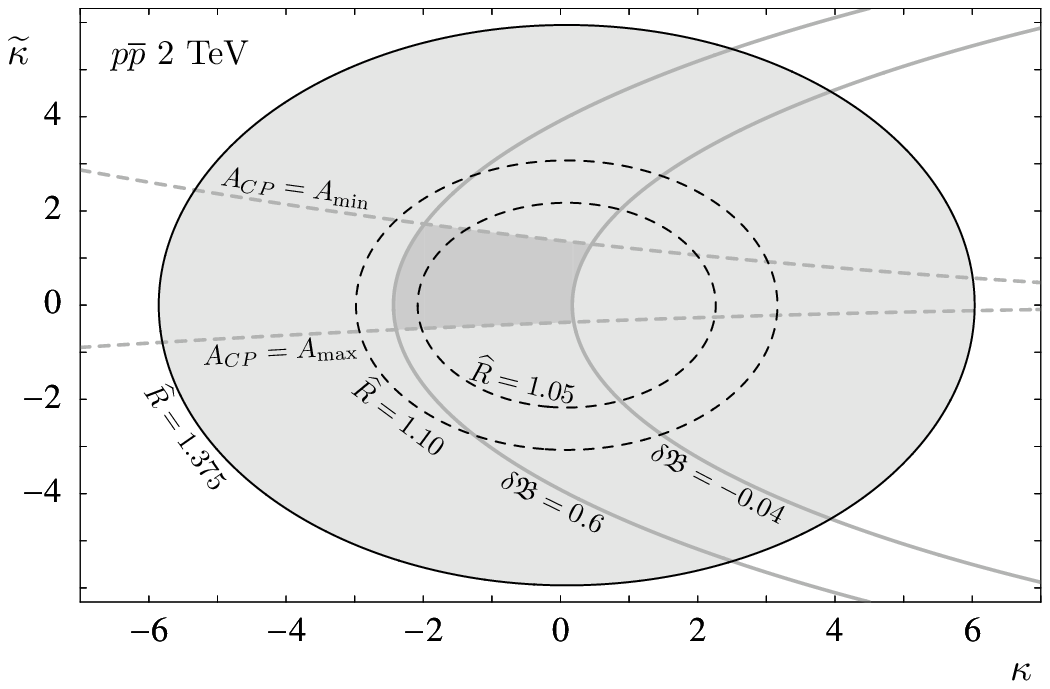}}
  \caption{The allowed parameter space for the anomalous magnetic and
    electric dipole moments of the top quark.  Gray solid lines:
    region allowed by the experimental results for the branching ratio
    for $B\to X_s \gamma$, see eq.\ (\ref{brnum}).  Gray dashed lines:
    region allowed by the experimental results for the $CP$ asymmetry
    for $B\to X_s \gamma$, see eq.\ (\ref{eq:acpineq}).  Black solid line:
    region allowed by the CDF measurement of $\widehat{R}$ for
    $t\overline{t}\gamma$ production at $\sqrt{s}=2$ TeV, see eq.\
    (\ref{tevatronratio}), with the cuts (\ref{eq:cuts}),
    (\ref{eq:cuts2}).  Black dashed lines: regions allowed by the
    values of $\widehat{R}$ indicated in the figure.  }
  \label{fig:fig1}
\end{figure}

\begin{figure}[p]
  \centering
  \scalebox{1.0}{\includegraphics{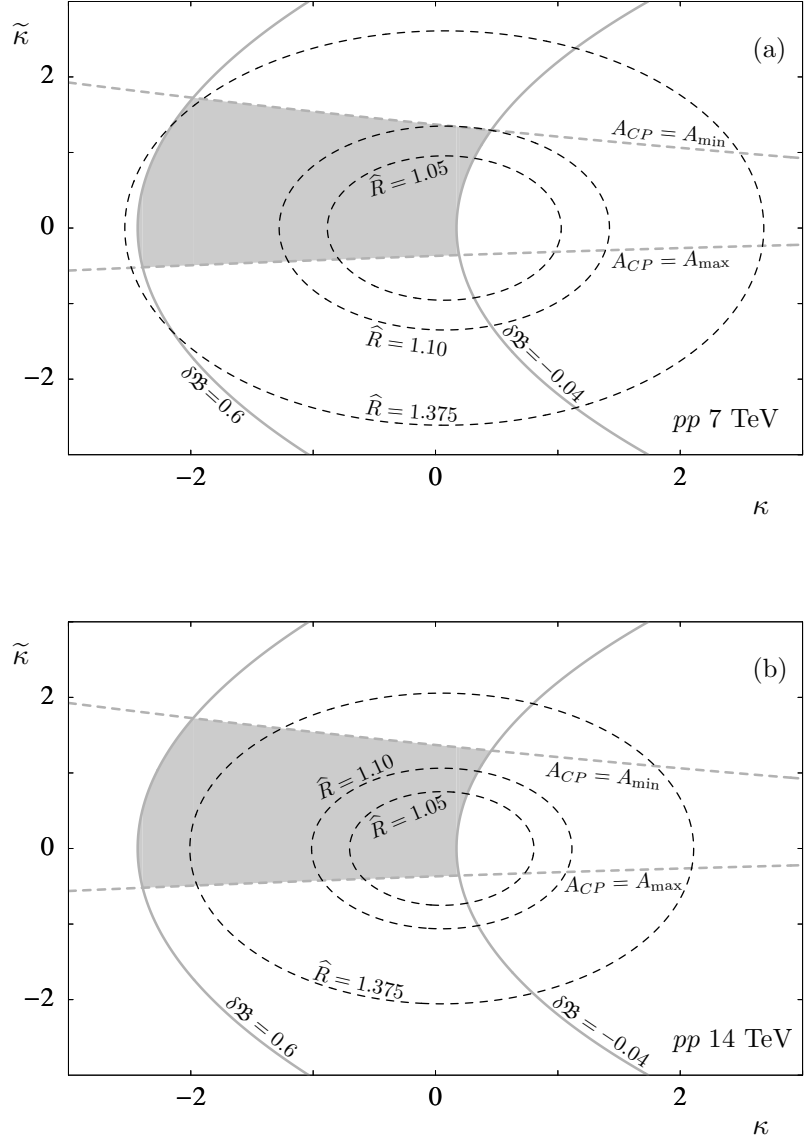}}
  \caption{Gray lines as in previous figure.  Black solid and dashed
    lines delimit the regions allowed by hypothetical measurements of
    $\widehat{R}$ for $t\overline{t}\gamma$ production with the cuts
    (\ref{eq:cuts}), (\ref{eq:cuts2}) at the LHC at (a) $\sqrt{s}=7$
    TeV, (b) $\sqrt{s}=14$ TeV.}
  \label{fig:fig2}
\end{figure}

We can now use (\ref{tevatronratio}), (\ref{brnum}) and
(\ref{means}) to constrain the allowed region in the $\kappa$ vs.
$\widetilde{\kappa}$ plane.  For the branching ratio, the region
allowed at the 1$\sigma$ level is seen from (\ref{brnum}) to be
bounded by 
\begin{equation}
  \label{eq:brnchineq}
  -0.0383<\delta{\cal B}(B\to X_s \gamma)<0.5983.
\end{equation}
That region is delimited in figures \ref{fig:fig1}, \ref{fig:fig2} by
gray solid lines.  Roughly speaking the MDM is bounded to be $-2 <
\kappa < 1$ which translated to the $m_t \mu_t = \kappa e/2 = 0.15
\kappa$ term used in \cite{kamenik} means that $-0.3 < m_t \mu_t <
0.15$.  Our limits are significantly more stringent than reported in
\cite{kamenik}. 

With the experimental value for $A_{\rm CP}$ quoted above and its
expression (\ref{means}), at the 1$\sigma$ level the asymmetry must
satisfy the inequalities
\begin{equation}
  \label{eq:acpineq}
-0.8 - \sqrt{2.9^2+\delta A_{\rm CP}[\%]^2}
< \overline{A}_{\rm CP}[\%]
<  -0.8 + \sqrt{2.9^2+\delta A_{\rm
    CP}[\%]^2}, 
\end{equation}
which define the region in the $\kappa,\widetilde{\kappa}$ plane
allowed by the measured asymmetry.  That region is shown in figures
\ref{fig:fig1}, \ref{fig:fig2} by gray dashed lines, with the shaded
area corresponding to the region allowed by both  measurements, ${\cal
  B}$ and $A_{\rm CP}$. 

The measurement of $\widehat{R}$ at the Tevatron by the CDF
collaboration sets limits on $(\kappa,\widetilde{\kappa})$ through
(\ref{tevatronratio}).  At the 1$\sigma$ level the allowed region for
$(\kappa,\widetilde{\kappa})$ is bounded by the inequalities $0.625 <
\widehat{R} < 1.375$.  The lower value turns out to be unattainable,
so it does not set any bound.  The region delimited by
$\widehat{R}=1.375$ is shown in figure \ref{fig:fig1} by the black
solid line.  The black dashed lines in that figure show the regions
that would be delimited by hypothetical measurements $\widehat{R}=1\pm
0.1$ and $1\pm 0.05$.  We see from the figure that, as expected from
the analysis in \cite{baur04}, the bounds set by the Tevatron
measurement of $\widehat{R}$ are much less constraining than those
arising from the asymmetry and branching ratio for $B\to X_s \gamma$.
This is so even in the hypothetical case of an experimental result
$\widehat{R}=1\pm 0.1$ with a 10\% measurement error.  Only a 5\%
measurement uncertainty could yield bounds of the same order of
magnitude at most. 

\begin{figure}[p]
  \centering
  \scalebox{1.0}{\includegraphics{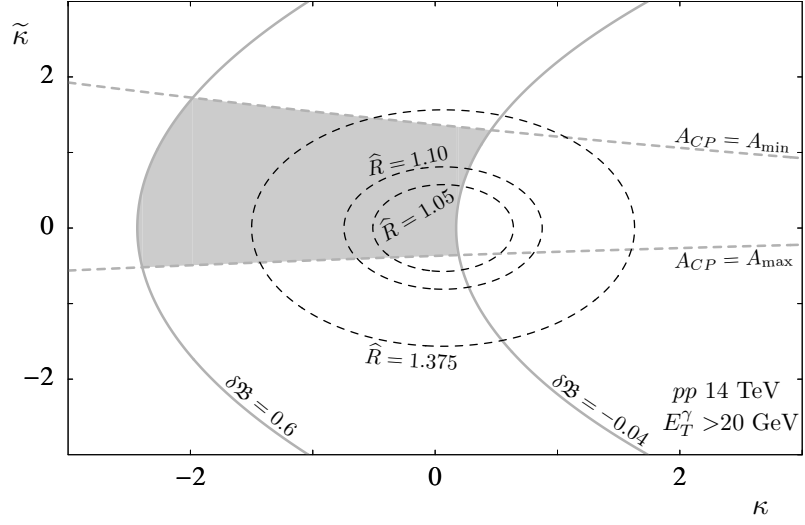}}
  \caption{Same as figure \ref{fig:fig2}(b), but with the stricter cut 
    $E_T^\gamma>20$ GeV.}
  \label{fig:fig3}
\end{figure}

We have also performed the same analysis for hypothetical measurements
of $\widehat{R}$ in $pp$ collisions at the LHC, with the same
semileptonic final states and cuts (\ref{eq:cuts}), (\ref{eq:cuts2}).
The results are shown in figure \ref{fig:fig2} (a) for the lower LHC
energy and in figure \ref{fig:fig2} (b) for the higher one.  As seen
in the figure, the hypothetical experimental results at the LHC would
remove significant portions of the region of the
$(\kappa,\widetilde{\kappa})$ plane allowed by the measurements of the
branching ratio and $CP$ asymmetry of $B\to X_s \gamma$.  Whereas this
is true already at $\sqrt{s}=7$ TeV, the constraints set by a
measurement of $\widehat{R}$ at $\sqrt{s}=14$ TeV with an experimental
uncertainty smaller than, say, 30\% would lead to remarkably tighter
bounds on $(\kappa,\widetilde{\kappa})$ than those currently
available.  We remark here that the cuts we have applied are rather
conservative.  Indeed, due to the higher cross sections at LHC
collision energies, and to the LHC high luminosity, more stringent
cuts could be enforced that could significantly improve the
sensitivity of $t\overline{t}\gamma$ production to top dipole moments
while still yielding high enough statistics.  As a simple illustration
of this, we show in figure \ref{fig:fig3} the bounds that would be
obtained at $\sqrt{s}=$ 14 TeV if in (\ref{eq:cuts}), (\ref{eq:cuts2})
we substitute the cut $E_T^\gamma > 10$ GeV by $E_T^\gamma > 20$ GeV.
As a result of that stricter cut the cross section decreases from
$\sigma_{t\overline t\gamma}^\mathrm{SM}=$ 0.8034 pb to 0.4577 pb,
which is still almost 25 times larger than the corresponding cross
section at the Tevatron. As seen in figure \ref{fig:fig3}, the
sensitivity is increased with respect to figure \ref{fig:fig2} (b) by
30\%.  Whereas the parton-level analysis carried out here is not the
appropriate context to discuss the optimization of experimental cuts,
we believe that our results demonstrate the interest of such detailed
studies.

On the other hand, the semileptonic channel considered here by analogy
with the CDF measurement \cite{cdfmeasurement} may not necessarily be
the only experimentally relevant one.  The question then arises how
robust our estimates of the sensitivity to the top dipole moments of
$t\overline{t}\gamma$ production are with respect to variations of the
selected final state.  As a rough attempt to an answer we have
considered the process $p\overline{p}$ or $pp\rightarrow
t\overline{t}\rightarrow b\overline{b}W^+W^-\gamma$, with only the cut
$E_T^\gamma>10$ GeV, for which we performed the same analysis as
described above.  In this case we carried out the computations with
\textsc{CalcHep} 3.4 \cite{pukhov}. Besides the expected numerical
differences in the results, the conclusions drawn from that alternate
analysis are fully consistent with those obtained from the more
detailed study presented here.

\section{Conclusions}
\label{sec:conclusions}

We have discussed in the foregoing sections the bounds on the top
anomalous dipole moments that can be obtained from measurements of the
semi-inclusive decays $B\rightarrow X_s\gamma$, and of
$t\overline{t}\gamma$ production at the Tevatron and the LHC.  We
reviewed the experimental and theoretical determinations of the
branching fraction and $CP$ asymmetry of $B\rightarrow X_s\gamma$ and
obtained from them bounds on the top MDM and EDM that are
significantly more stringent than those reported in the previous
literature.  The allowed region is defined by the shaded area
in the $(\kappa,\widetilde{\kappa})$ plane as shown in
the figures.  Roughly speaking, the MDM term is bounded
by $-2 < \kappa < 0.3$ whereas the EDM term is bounded by
$-0.5 < \widetilde \kappa < 1.5$.  We can translate these limits
in terms of the well known MDM factor $(g-2)/2 = a_t =
3/2\,\kappa$:  $-3 < a_t < 0.45$ and the EDM factor
$d_t = 0.57 \times 10^{-16} \widetilde \kappa$:
$-0.29< d_t < 0.86 \times 10^{-16} \; e\, $ cm.

We carried out a detailed LO computation of $t\overline{t}\gamma$
production at the Tevatron and the LHC, from which we extracted bounds
on the anomalous top MDM and EDM that we compare to those coming from 
$B\rightarrow X_s\gamma$.  From that comparison we conclude that the
bounds obtained from the measurement \cite{cdfmeasurement} at the
Tevatron are too weak to be relevant, but similar studies at the LHC
could significantly improve the bounds from $B\rightarrow X_s\gamma$.
This conclusion confirms a previous assessment in \cite{baur04} using
a different approach.

The estimates presented in this paper of the direct bounds on the top
MDM and EDM that could be obtained from $t\overline{t}\gamma$
production at the LHC, especially at 14 TeV, remove large portions of
the parameter space allowed by the indirect bounds from $B\rightarrow
X_s\gamma$.  Thus, the combination of both sets of bounds could lead
to strikingly tighter bounds on $(\kappa,\widetilde{\kappa})$ than
those coming from $B\rightarrow X_s\gamma$ alone.


\noindent
{\bf Acknowledgments}~~~
We thank Conacyt and SNI for support.



\end{document}